\documentclass{article}

\usepackage[nonatbib,final]{neurips_2020_ml4ps}

\usepackage[utf8]{inputenc} %
\usepackage[T1]{fontenc}    %
\usepackage{hyperref}       %
\usepackage{url}            %
\usepackage{booktabs}       %
\usepackage{amsfonts}       %
\usepackage{nicefrac}       %
\usepackage{microtype}      %
\usepackage{graphicx}
\usepackage{subcaption}

\usepackage[table]{xcolor} %
\usepackage{amsmath}

\newcommand{\eg}{\textit{e.g.}}
\newcommand{\etc}{\textit{etc}}

\DeclareMathOperator*{\argmin}{arg\,min}
\newcommand{\angstrom}{\mbox{\normalfont\AA}}

\title{Solving Inverse Problems for Spectral Energy Distributions with Deep Generative Networks}

\author{%
  Agapi Rissaki\\
  National Technical University of Athens\\
  \texttt{arissaki@corelab.ntua.gr} \\
  \And
  Orestis Pavlou\\
  European University Cyprus\\
  \texttt{o.pavlou@euc.ac.cy} \\
  \And
  Dimitris Fotakis\\
  National Technical University of Athens\\
  \texttt{fotakis@cs.ntua.gr} \\
  \And
  Vicky Papadopoulou\\
  European University Cyprus\\
  \texttt{v.papadopoulou@euc.ac.cy} \\
  \And
  Andreas Efstathiou\\
  European University Cyprus\\
  \texttt{a.efstathiou@euc.ac.cy} \\
}

\begin{document}

\maketitle

\begin{abstract}
  We propose an end-to-end approach for solving inverse problems for a class of complex astronomical signals, namely Spectral Energy Distributions (SEDs). Our goal is to reconstruct such signals from scarce and/or unreliable measurements. We achieve that by leveraging a learned structural prior in the form of a Deep Generative Network. Similar methods have been tested almost exclusively for images which display useful properties (e.g., locality, periodicity) that are implicitly exploited. However, SEDs lack such properties which make the problem more challenging. We manage to successfully extend the methods to SEDs using a Generative Latent Optimization model trained with significantly fewer and corrupted data.
\end{abstract}

\section{Introduction}
\label{intro}

In astrophysics, distributions constructed by energy measurements in different wavelengths, namely Spectral Energy Distributions (SEDs), are important tools for studying the physical properties and evolution of astronomical objects. SEDs can be used for example to determine the luminosity of astronomical objects, the rate at which galaxies form new stars or the rate at which supermassive black holes accrete mass to generate energy in quasars \cite{Efstathiou1, Efstathiou2, rowanrobinson}. However, the measurement process is prone to statistical (random) as well as systematic errors, such as background and foreground noise interference, \textit{i.e.}, atmospheric absorption and distortion, opaque/obscuring dust, \etc. Due to these factors, as well as technical limitations, such as camera sensor sensitivity, cooling, resolution, \etc, SEDs are collected in scarce, often incomplete datasets. SEDs are compared to physical models in order to find the best-fit model(s), which provides an insight into the underlying physical processes and properties of the target. This highlights the importance of expanding the range and improving the accuracy of the available data points. In the literature, computational methods have been widely used to enhance SEDs and handle the experimental error \cite{Walcher_2010}. 

In recent years, deep learning has proven to be an important tool for enhancement of real data and in general for solving inverse problems, where the goal is to reconstruct or correct a signal given an incomplete and/or noisy version. Specifically for astronomical data, deep learning techniques have been used mainly for astronomical imaging, such as deblending images of galaxies \cite{Boucaud_2019} or image enhancement \cite{lanusse2019hybrid}. For SEDs, deep learning has been used in forward problems such as feature extraction \cite{refId0}, but not inverse problems. In this paper we use well-known deep learning techniques adjusted appropriately in order to solve various inverse problems for SEDs. 

The method we apply is data-driven and utilizes Deep Generative Models as learned structural priors. More specifically, models like Variational AutoEncoders (VAEs) \cite{vae1} and Generative Adversarial Networks (GANs) \cite{gans}, trained on large datasets (most frequently of images) are able to extract information about the underlying data distribution and generate realistic samples. These models, once trained, can be used as structural priors for solving inverse problems \cite{bora17}. Thus, this methods requires us to train a high-quality generative network which can model realistic SEDs, with properties such as high-frequency, irregularity \etc. In this paper, we use the Generative Latent Optimization framework (GLO) \cite{glo} to train a deep generative network suitable for our needs. The framework allows us to train a high-quality generative network with more flexibility than a VAE and at the same time offers training efficiency unlike GANs, which are notoriously hard to train.

In order to train a generative network any state-of-the-art method requires a high-quality large dataset. However, for the case of SEDs these prerequisites are unrealistic since the measurement procedure contains innate error, incompleteness and is particularly expensive. To overcome the issue of erroneous and/or incomplete samples we propose an end-to-end approach: ($1$) a preprocessing step where we utilize classical computational methods for enhancement, \eg, iterative PCA \cite{Walcher_2010}, ($2$) the deep learning method described above. Our approach is useful for a variety of inverse problems and it can mitigate the long-term cost of solving such problems for SEDs. Furthermore, it is expected to improve overall performance on these problems even with significant corruption and/or incompleteness by leveraging the powerful generalization property as well as the robustness of a deep generative network.

\section{Method}

Suppose, we collect measurements of the form:
\begin{equation} \label{noisyCS}
y = Ax^* + \eta\,,
\end{equation} 
where $A$ is a measurement matrix and $\eta$ a noise vector. Our goal is to reconstruct the signal $x^*$, given $A$ and $\eta$, thus solve the linear inverse problem. This formulation usually refers to compressed sensing (where we assume few measurements taken) but can be also used to model several real-world problems concerning SEDs.%
We tackle the problem for the case of SEDs using a deep generative network as a structural prior \cite{bora17}, a method that has been successfully applied to natural images.

\subsection{Building the Generative Network}

To build our generative network we use the Generative Latent Optimization (GLO) framework \cite{glo}, which allows us to train a relatively large generator (sufficiently over-parametrized) in order to achieve good generalization \cite{generalization_DNN}. The framework is based on the manipulation of the generator's latent space as well as its parameters using a simple reconstruction loss. We use the GLO framework as an alternative to GANs which are trained via an adversarial optimization scheme. Unlike GLO, which consists of a simple loss minimization back-propagated to the latent space, GANs should ideally converge to an (approximate) equilibrium which is not guaranteed and/or requires excessive resources \cite{gan_equilibria}. Thus, when training GANs in practice it is common to examine the generated samples and stop the training when they are satisfactory. In the case of images this technique can be easily applied, but for SEDs this is not feasible. In fact, we use the ability to solve inverse problems as a proxy to evaluate our trained generator.

Let us examine the training procedure more closely. We train the generator $G \; : \; \mathcal{Z} \rightarrow \mathcal{X}$, where $\mathcal{Z}$ denotes the latent space and $\mathcal{X}$ the underlying class of SEDs which is described by the training set $\{ x_i \}_{i=1}^N$. Prior to training, we randomly initialize the latent codes $z_i \in \mathcal{Z}$ from a multi-dimensional Normal distribution and pair them with each of the samples $x_i$. During training, the generator's parameters and the latent codes $\{ z_i \}_{i=1}^N$ are jointly optimized, as described by \eqref{glo_training}. The optimization is driven by a simple reconstruction loss $\mathcal{L}(\cdot)$, which in our case is Mean Squared Error (MSE). 

\begin{equation} \label{glo_training}
\min\limits_{G} \frac{1}{N} \sum_{i=1}^N \min\limits_{z_i \in \mathcal{Z}} \left[ \mathcal{L}(G(z_i), x_i) \right]
\end{equation}

More specifically, the gradient of the loss function with respect to the parameters of the generator and the latent code is back-propagated all the way through the network and to the latent space. This training procedure makes the latent space more structurally meaningful and suitable for reconstruction. To promote this feature, we project the latent codes onto the unit sphere during training \cite{glo}.

\subsection{Reconstruction}

Given the generative network $G(\cdot)$, the estimated solution of an inverse problem \eqref{noisyCS} could be $\hat{x} = G(\hat{z})$ where:
\begin{equation} \label{recon1}
\hat{z} = \argmin\limits_{z\in \mathcal{Z}} \frac{1}{m} ||AG(z) - y||^2_2
\end{equation} 
In other words, we (approximately) optimize the latent code $\hat{z}$ such that the corresponding signal $\hat{x}$ matches the measurements $y$. We optimize $\hat{z}$ by back-propagating the gradient of the reconstruction loss through $G(\cdot)$ \cite{bora17}. Note that we have to project $z$ onto the unit sphere, similarly to training. 
In a different approach \cite{bora17}, instead of explicitly projecting $z$ onto the unit sphere, we can apply a regularization to implicitly restrict $z$ as follows:
\begin{equation} \label{recon2}
\hat{z} = \argmin\limits_{z\in \mathcal{Z}} \frac{1}{m} ||AG(z) - y||^2_2 + R(z)\,,
\end{equation}
where $R(z) = \lambda ||z||^2_2$ is the regularization term and $\lambda$ a balance hyperparameter.

\section{Experiments}

\subsection{Data}

We apply our approach to Sloan Digital Sky Survey (SDSS) spectra \footnote{https://www.sdss.org/dr12/spectro/}. Specifically, we use the preprocessed SDSS Corrected Spectra dataset offered by the astroML library \cite{astroml}. The dataset contains SEDs for $4000$ galaxies moved to restframe, preprocessed with iterative PCA and resampled to $1000$ wavelengths ($3000-8000 \angstrom$). Although the preprocessing is imperfect, leading to outlier values, our deep learning approach still displays great performance due to its robustness. Notice that the original SDSS dataset consists of innately incomplete and/or corrupted SEDs, due to the nature of the measurement process. Thus, the original SEDs cannot be used directly for evaluation purposes because we would lack the ground truth. Instead, we consider part of the corrected SEDs produced by the preprocessing step as test data ($10\%$ of the preprocessed dataset), which we use for comparisons in Section \ref{results}.

\subsection{Training}

We train a Feed-forward neural network with $7$ hidden layers and leakyReLU activations (except for the output layer). We use $90\%$ of the preprocessed dataset as our training data and train our network for $10000$ epochs with batches of $64$ spectra. We use Adam optimization \cite{adam} with learning rate $0.1$ for the network's parameters and $0.01$ for the latent codes as well as 1d-batch normalization to accelerate the training procedure \cite{batchnorm}. We choose a simple Mean Squared
Error (MSE) as our loss function and we also apply weight decay to avoid overfitting. 

Our spectra consist of measurements for $1000$ wavelengths. We choose $50$ dimensions for the latent space, which are sufficient for the representation and allow for efficient training and reconstruction. For the reconstruction, we limit the optimization procedure to $1000$ epochs and choose a configuration similar to training. The project is developed using PyTorch \cite{pytorch}.

\subsection{Results} \label{results}

\begin{figure*}[t]
    \centering
    
    \begin{subfigure}{\linewidth}
        \centering
        \includegraphics[width=0.5\linewidth]{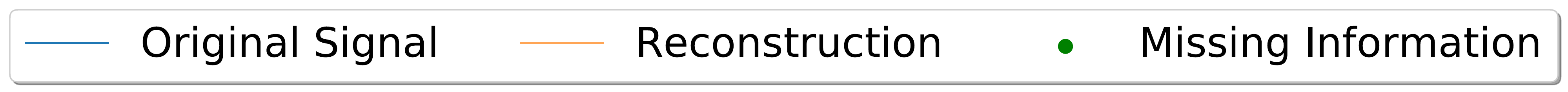}
    \end{subfigure}
    
    \begin{subfigure}{\linewidth}
        \centering
        \includegraphics[width=1\linewidth]{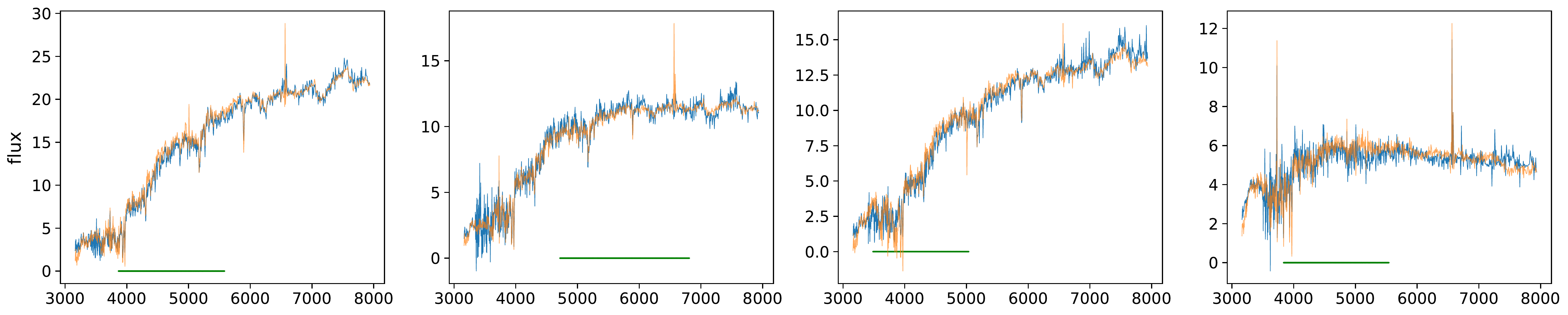}
        \caption{Inpainting problem: measurements are missing in a continuous window.}
		\label{a}
    \end{subfigure}
    \vspace{-0.3mm}

    \begin{subfigure}{\linewidth}
        \centering
        \includegraphics[width=1\linewidth]{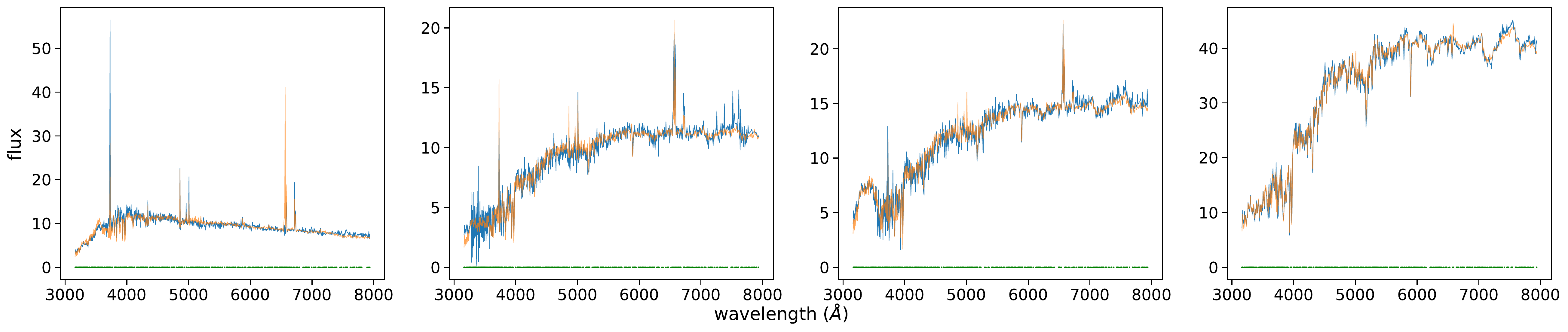}
        \caption{Super-resolution problem: randomly selected measurements are missing.}
		\label{b}
    \end{subfigure}
    \vspace{-0.3mm}
    
    \caption{The original SED signals and their reconstruction for $40\%$ missing information in inpainting (top) and super-resolution (bottom) settings.}
	\label{qual_samples}
\end{figure*}

\begin{figure*}

\centering
\subfloat[]{\label{fig:inpainting_quant}
\centering
\includegraphics[width=0.42\linewidth]{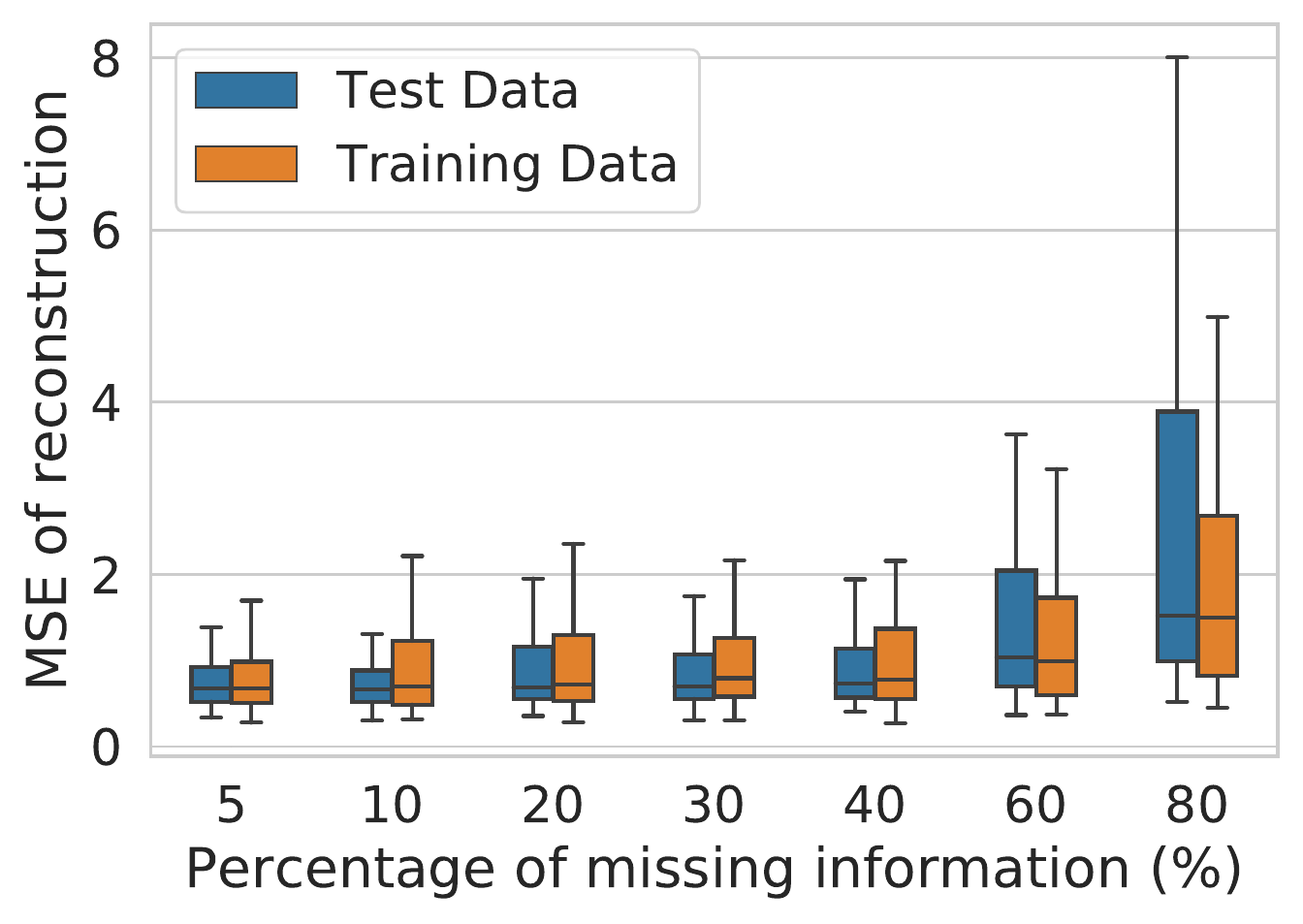}
}
\subfloat[]{\label{fig:denoising_quant}
\centering

\includegraphics[width=0.42\linewidth]{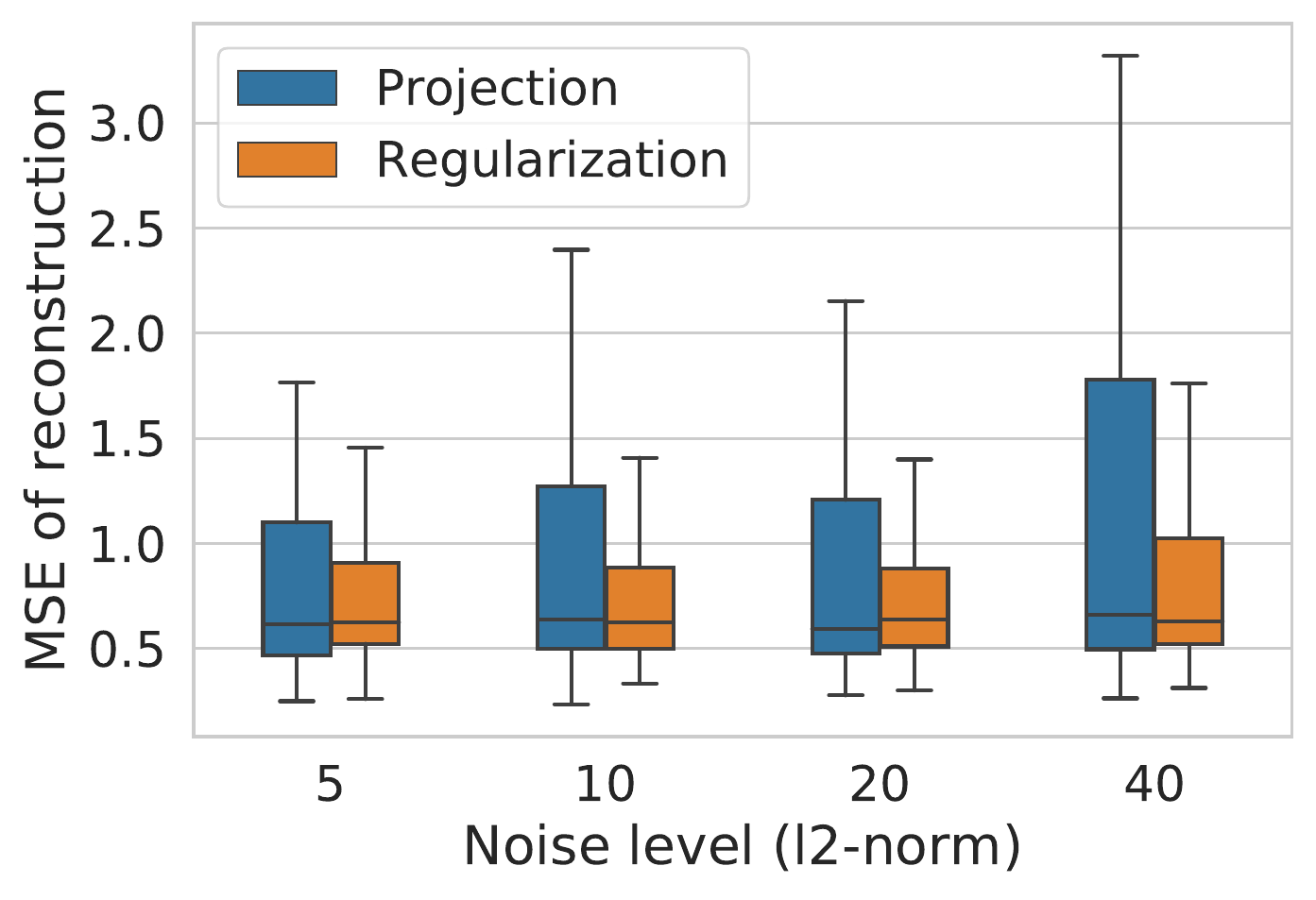}
}

\caption{MSE of reconstruction for $100$ randomly selected signals on: (a)  inpainting, for different levels of missing information (\%). (b) denoising, for different levels of added noise ($l_2$-norm).}
\label{quant}
\end{figure*}

We evaluate our approach, both qualitatively and quantitatively, for different inverse problems, by artificially injecting realistic corruption and/or incompleteness to our test data. For the qualitative evaluation (Figure \ref{qual_samples}), we examine the performance of our algorithm on inverse problems with missing information. More specifically, the missing information corresponds to either a continuous window of missing values (inpainting) or randomly chosen values throughout the entire signal (super-resolution). For each problem we randomly select four SED signals from our test set, then apply the appropriate masking and produce a reconstruction. The missing information in both cases is chosen to be $40\%$ of the total number of measurements that compose each SED. We can see that for both problems, the reconstruction closely follows the trajectory of the original signal and in most cases predicts the high-frequency changes and large spikes. 

In Figure \ref{quant}, we examine quantitatively the performance of our algorithm on the problems of (a) inpainting and (b) denoising (that is removing added noise drawn by a normal distribution). For each configuration, we show the reconstruction MSE of $100$ randomly selected SEDs (excluding measurements that fall outside $1.5$ times the interquartile range). 
In Figure \ref{quant}(a), we examine the performance for different levels of missing information and compare between SEDs drawn from test and training data. In both cases the MSE of the vast majority of signals is particularly low for up to $40\%$ of missing information. For reasonable percentages of missing information the performance on test data is on par with the training data. For a sufficient proportion of the examined signals, this trend persists even for larger percentages (see median values). Given that our generative network was optimized to represent the training data, this shows a considerable generalization capability, which is crucial for the effectiveness of our approach. In Figure \ref{quant}(b), we examine the performance for different levels of added noise and compare between our reconstruction methods, the explicit projection (eq. \ref{recon1}) and the regularization (eq. \ref{recon2}). We can see that for all levels of added noise the MSE is particularly low, which indicates notable performance on denoising. Furthermore, when regularization is utilized we observe better error concentration, which can be attributed to the flexibility it offers to the reconstruction process.

\section{Conclusion and Future Work}

We presented an end-to-end deep learning solution for various inverse problems concerning Spectral Energy Distributions (SEDs). Our approach relies on a deep generative network, tailored to the particular properties of SEDs, as a structural prior leveraging its generalization capability. Our preliminary results show promising performance on realistic inverse problems. We are working to extend this project to diverse and more demanding SED families \eg, for different parts of the spectrum. Another future direction involves transfer learning techniques, as well as ensemble learning in order to extend our approach to data that are even more incomplete. Finally, we could augment our method using bi-directional training in order to simultaneously extract information regarding the astrophysical objects we study. This idea draws from recent research on invertible neural networks for inverse problems \cite{invertibleNNs}.

\section*{Broader Impact}

This project will have broad impact in the effort to interpret the SEDs which will be made available with a number of current and future ground-based and space missions such as LSST, Euclid, JWST and SPICA. Although the examples used in this work concentrate on the optical part of the spectrum, the same method can also be used on SEDs which cover the whole spectrum of galaxies from the ultraviolet to the radio. Such studies of the complete SEDs of galaxies are now recognized as essential for a complete understanding of the processes that control galaxy formation and evolution (e.g. \cite{rowanrobinson, shirley}).

\bibliographystyle{unsrt}
\bibliography{bibliography}

\begin{thebibliography}{10}

\bibitem{Efstathiou1}
Andreas Efstathiou and Michael Rowan-Robinson.
\newblock Dusty discs ih active galactic nuclei.
\newblock {\em Monthly Notices of the Royal Astronomical Society},
  273(3):649--661, 1995.

\bibitem{Efstathiou2}
Andreas Efstathiou, Michael Rowan-Robinson, and Ralf Siebenmorgen.
\newblock Massive star formation in galaxies: radiative transfer models of the
  uv to millimetre emission of starburst galaxies.
\newblock {\em Monthly Notices of the Royal Astronomical Society},
  313(4):734–744, 2000.

\bibitem{rowanrobinson}
Michael Rowan-Robinson and et~al.
\newblock Spectral energy distributions and luminosities of galaxies and active
  galactic nuclei in the spitzer wide-area infrared extragalactic (swire)
  legacy survey.
\newblock {\em The Astronomical Journal}, 129:1183--1197, March 2005.

\bibitem{Walcher_2010}
Jakob Walcher, Brent Groves, Tamás Budavári, and Daniel Dale.
\newblock Fitting the integrated spectral energy distributions of galaxies.
\newblock {\em Astrophysics and Space Science}, 331(1):1–51, Aug 2010.

\bibitem{Boucaud_2019}
Alexandre Boucaud, Marc Huertas-Company, Caroline Heneka, Emille E~O Ishida,
  Nima Sedaghat, Rafael~S de~Souza, Ben Moews, Hervé Dole, Marco Castellano,
  Emiliano Merlin, and et~al.
\newblock Photometry of high-redshift blended galaxies using deep learning.
\newblock {\em Monthly Notices of the Royal Astronomical Society},
  491(2):2481–2495, Dec 2019.

\bibitem{lanusse2019hybrid}
Francois Lanusse, Peter Melchior, and Fred Moolekamp.
\newblock Hybrid physical-deep learning model for astronomical inverse
  problems, 2019.

\bibitem{refId0}
{Frontera-Pons, J.}, {Sureau, F.}, {Bobin, J.}, and {Le Floc\'{}h, E.}
\newblock Unsupervised feature-learning for galaxy seds with denoising
  autoencoders.
\newblock {\em A\&A}, 603:A60, 2017.

\bibitem{vae1}
Diederik~P. Kingma and Max Welling.
\newblock An introduction to variational autoencoders.
\newblock {\em CoRR}, abs/1906.02691, 2019.

\bibitem{gans}
Ian~J. Goodfellow, Jean Pouget-Abadie, Mehdi Mirza, Bing Xu, David
  Warde-Farley, Sherjil Ozair, Aaron Courville, and Yoshua Bengio.
\newblock Generative adversarial networks, 2014.

\bibitem{bora17}
Ashish Bora, Ajil Jalal, Eric Price, and Alexandros~G. Dimakis.
\newblock Compressed sensing using generative models.
\newblock In {\em Proceedings of the 34th International Conference on Machine
  Learning - Volume 70}, ICML'17, pages 537--546. JMLR.org, 2017.

\bibitem{glo}
Piotr Bojanowski, Armand Joulin, David Lopez-Paz, and Arthur Szlam.
\newblock Optimizing the latent space of generative networks.
\newblock 2017.

\bibitem{generalization_DNN}
Behnam Neyshabur, Srinadh Bhojanapalli, David Mcallester, and Nati Srebro.
\newblock Exploring generalization in deep learning.
\newblock In I.~Guyon, U.~V. Luxburg, S.~Bengio, H.~Wallach, R.~Fergus,
  S.~Vishwanathan, and R.~Garnett, editors, {\em Advances in Neural Information
  Processing Systems 30}, pages 5947--5956. Curran Associates, Inc., 2017.

\bibitem{gan_equilibria}
Frans~A. Oliehoek, Rahul Savani, Jose Gallego{-}Posada, Elise van~der Pol, and
  Roderich Gro{\ss}.
\newblock Beyond local nash equilibria for adversarial networks.
\newblock {\em CoRR}, abs/1806.07268, 2018.

\bibitem{astroml}
J.T. {Vanderplas}, A.J. {Connolly}, {\v Z}.~{Ivezi{\'c}}, and A.~{Gray}.
\newblock Introduction to astroml: Machine learning for astrophysics.
\newblock In {\em Conference on Intelligent Data Understanding (CIDU)}, pages
  47--54, Oct 2012.

\bibitem{adam}
Diederik Kingma and Jimmy Ba.
\newblock Adam: A method for stochastic optimization.
\newblock {\em International Conference on Learning Representations}, 12 2014.

\bibitem{batchnorm}
Sergey Ioffe and Christian Szegedy.
\newblock Batch normalization: Accelerating deep network training by reducing
  internal covariate shift.
\newblock In {\em Proceedings of the 32nd International Conference on
  International Conference on Machine Learning - Volume 37}, ICML’15, page
  448–456. JMLR.org, 2015.

\bibitem{pytorch}
Adam Paszke, Sam Gross, Soumith Chintala, Gregory Chanan, Edward Yang, Zachary
  DeVito, Zeming Lin, Alban Desmaison, Luca Antiga, and Adam Lerer.
\newblock Automatic differentiation in pytorch.
\newblock 2017.

\bibitem{invertibleNNs}
Lynton Ardizzone, Jakob Kruse, Sebastian~J. Wirkert, Daniel Rahner, Eric~W.
  Pellegrini, Ralf~S. Klessen, Lena Maier{-}Hein, Carsten Rother, and Ullrich
  K{\"{o}}the.
\newblock Analyzing inverse problems with invertible neural networks.
\newblock {\em CoRR}, abs/1808.04730, 2018.

\bibitem{shirley}
Raphael {Shirley}, Yannick {Roehlly}, Peter~D. {Hurley}, Veronique {Buat},
  Mar{\'\i}a del~Carmen {Campos Varillas}, Steven {Duivenvoorden}, Kenneth~J.
  {Duncan}, Andreas {Efstathiou}, Duncan {Farrah}, Eduardo {Gonz{\'a}lez
  Solares}, Katarzyna {Malek}, Lucia {Marchetti}, Ian {McCheyne}, Andreas
  {Papadopoulos}, Estelle {Pons}, Roberto {Scipioni}, Mattia {Vaccari}, and Seb
  {Oliver}.
\newblock {HELP: a catalogue of 170 million objects, selected at 0.36-4.5
  {\ensuremath{\mu}}m, from 1270 deg$^{2}$ of prime extragalactic fields}.
\newblock {\em Monthly Notices of the Royal Astronomical Society},
  490(1):634--656, November 2019.

\end{thebibliography}

\end{document}